\documentstyle[prl,aps,psfig]{revtex}

\begin{document}
\draft
\title{Effects of construction history on the stress distribution under a sand
pile}
\author{Loic Vanel$^{2}$, Daniel Howell$^{1}$, D. Clark$^{1},$R.~P. Behringer$^{1}$, and Eric
Cl\'{e}ment$^{2}$}
\address{1. Department of\\
Physics and Center for Nonlinear and Complex Systems, Duke University,\\
Durham NC, 27708-0305, USA}
\address{2. Laboratoire des Milieux D\'{e}sordonn\'{e}s et H\'{e}t\'{e}rog\`{e}nes,\\
UMR 7603, Case 86, Universit\'{e} Piere et Marie Curie, 4, Place Jussieu,\\
75252 Paris Cedex, France}
\date{\today}
\maketitle

\begin{abstract}
We report experiments on cohesionless granular piles to determine the
effect of construction history on the static stress distribution. The
stresses beneath the piles are monitored using a very sensitive
capacitive technique.  The piles are formed either by release of
granular material from a relatively small output (localized source),
or from a large diameter sieve (homogeneous rain). The stress profiles
resulting from localized source inputs have a clear stress dip near
the center of the pile while the results from an homogeneous rain show
no stress dip. We also show that the stress profiles scale simply with
the pile height. Experiments on wedges-shaped piles show the same
effects but to a lesser degree.
\end{abstract}


\twocolumn

\begin{figure}[]
\psfig{file=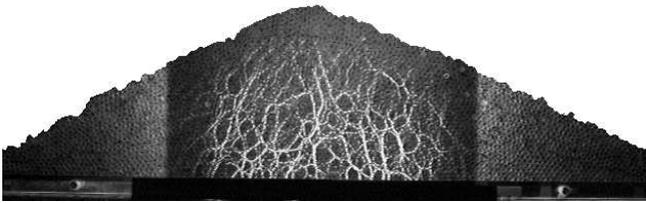,width=3.375in}
\caption{Two-dimensional pile of photoelastic disks created by a
localized-source procedure. The center section of the image is viewed
between crossed polarizers allowing one to see the underlying stress
structure.}
\label{fig:chains}
\end{figure}

Granular systems have captured much recent interest because of their
rich phenomenology, and important applications\cite{reviews}. Static
arrays show inhomogeneous spatial stress profiles called stress
chains\cite{chains}, where forces are carried primarily by a small
fraction of the total number of grains. Recent numerical
simulations\cite{radjai} and experiments\cite {howell} have shown that
the structure and the nature of these chains plays a critical role in
the dynamics and statics of dense granular systems even in
the absence of strong disorder of the granular packings \cite
{ouagenouni,eloy} (see Fig.~\ref{fig:chains}). Necessarily, the
presence of these chains must be reflected in the continuum
constitutive relations which are needed to close the governing
equations and thereby, solve even the simplest boundary value problems
in granular statics \cite
{bouchaud,watson,wittmer,wittmer2,savage,goddard}.

The stress profile under a static pile of granular material provides a
useful method for probing the effects of stress chains and the history
of their formation.  The literature contains many experiments
examining stress profiles under static piles of granular
material\cite{savage}.  Although there are a number of such studies,
they are not in mutual agreement.  In addition, a number of competing
constitutive models have been invoked to explain the experimental
observations.  

The present experiments have been carried out with the aim of
resolving the experimental conflict by determining as carefully as
possible the relation between the preparation of a heap and the stress
profile at its base.  It is important to clarify the effect of
construction technique on the stress structure for the following
reasons:\newline \indent 1. To help understand the wide variation in
past data.\newline \indent 2. To test some of the theories which
depend explicitly\newline \indent on construction
history\cite{wittmer}.

Of possible pile geometries, conical, and wedge-shaped heaps have been
the most frequently studied.  Many of the experiments on conical piles
have indicated, contrary to simple intuition, that there is a dip in
the pressure profile beneath the
center\cite{hummel,jotaki,smid,brockbank}. The existence of a dip in
the stress profile for wedge-shaped piles is an open
question\cite{savage,savage98}.

There are important technical considerations in determining whether
there is a stress dip.  The most important of these is the fact that
even modest deformations either of the surface supporting the pile or of the
force detector may lead to erroneous measurements\cite{savage}.  In
addition, if the pile is formed by dropping material onto the heap
from a considerable height, as opposed to gentler deposition methods,
it is likely that residual stresses become
frozen into the heap.  In such a case, or for a very heavy load,
there is likely to be a characteristic length associated with the
deformation of the pile under its own weight.    

There are only a few reported experiments addressing the influence of
the construction technique or filling rate on the stress.\cite{brown,huang,lee}
(However, two of us have probed the effect of the granular packing
history on the mean pressure at the bottom of a silo\cite{vanel}.)
Regarding sand-piles in particular, we are aware of only one set of
experiments that considered the effect of construction technique on
stress profiles, namely the work of Lee and Herrington for
wedge-shaped piles of sand\cite{lee}. These authors constructed piles
using three different methods and found that the different
construction techniques yielded results that were identical within the
resolution of their instruments; no dip was recorded.

In the present experiments, we explore the effects of construction
procedures on the pressure profiles using two different methods to
build both conical and wedge-shaped heaps.  We use detectors with
very high resolution and very small deflection.  We also build these
piles on rigid base plates. In the case of a conical pile, we
explicitly investigate the scaling of this profile with total mass for
piles with an height ratio up to three times that of the smallest
pile, corresponding to a mass ratio of $\sim 30$.

Several details of these experiments are important. We used sand of
diameter $1.2mm$ $\pm 0.4mm$ and angle of repose $32^{o}$. The base
plate on which we constructed most of these piles was a $15.0$ $mm$
thick duralumin support which was adequate to prevent deflection under
the weight of the pile. (Some additional experiments were carried out
using a $1.3mm$ steel base. These experiments used a fixed funnel
height, and they are discussed below.)  For a typical sand pile of
$H=8cm$ height, we estimated the maximal sagging of the bottom plate
to be $w_{m}=6.5\mu m$.  Therefore $w_{m}/H=10^{-5}$, a value that was
smaller by $\sim 10^{-3}$ than the relative deflection for which
sagging of the base might create a significant
perturbation\cite{savage98,lee,trollope}. A single capacitive normal
stress (i.e. pressure) sensor of diameter of $11.3$ mm ($9$ grain
diameters) was placed flush with the surface of the base plate.  We
then determined the normal stress at various locations along the
radial axis of the conical piles or along the short edge of the
wedge-shaped piles by repeated construction of heaps with the same
mass of sand. The resolution of the measuring device\cite{device} was
$0.25\%$ of the typical maximum stress for an $8cm$ pile, which
corresponded to a vertical deflection of the sensor of $\sim 1.3\mu
m$. We tested the consistency of the measurements with different
membrane thicknesses, and we found consistent results within
experimental resolution. Here, we present data obtained with only one
of these membranes which had a thickness $t=100\mu m$ . The sensor was
calibrated against the hydrostatic pressure of a water column.
However, the response of the sensor to known weights of granular
material was consistently somewhat smaller, by a factor of $\sim 0.9$,
than for water.  We emphasize that this reduction was constant
throughout the measurements.  In particular, using a calibration based
on granular mass, we consistently found that the integrated weight of
the pile was correct.

\begin{figure}[tbp]
\psfig{file=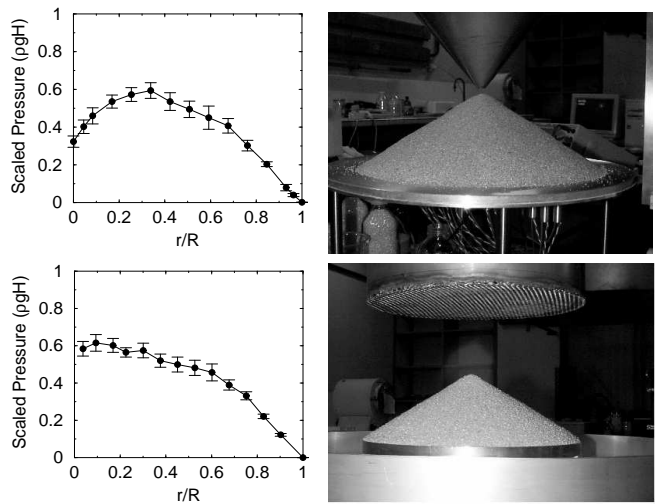,width=3.375in,angle=-90}
\caption{Normal stress profiles beneath conical piles of granular
materials of height $H$. The piles are made by different construction
techniques illustrated by the accompanying photographs (see text). The
distance from the center of the profiles is normalized by the pile
radius $R$ and stresses are normalized by $\rho gH$.}
\label{fig:cone}
\end{figure}

\begin{figure}[]
\psfig{file=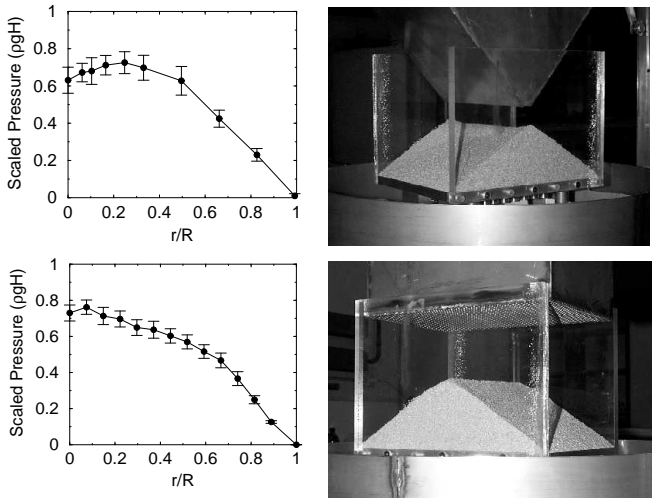,width=3.375in,angle=-90}
\caption{Normal stress profiles beneath wedge-shaped piles of granular
material of height $H$. The piles are made by different construction
techniques illustrated by the accompanying photographs (see text). The
distance from the center of the profiles is normalized by the pile
radius $R$ and stresses are normalized by $\rho gH$. }
\label{fig:wedge}
\end{figure}

We constructed both types of heaps by two qualitatively different
procedures.  The first of these used a funnel and we refer to it as
the `localized source' procedure; the second used a sieve, and we refer
to it as the `raining procedure'.  The following paragraphs give details
on each method.  Fig.~\ref{fig:cone} and Fig.~\ref{fig:wedge} show
photographs of these two configurations.

{\em The localized-source procedure:} We formed the pile using a
funnel with an outlet that was much smaller than the final pile
diameter. The funnel lifted steadily and slowly so that the outlet was
always slightly above the apex during the heap formation. This
approach, as opposed to a fixed funnel height, avoided effects from
the deposition of particles with large kinetic energies that varied
with the distance between the apex of the heap and the bottom of the
funnel\cite{hummel,jotaki}. For conical piles, the sand emptied from a
conical funnel of outlet diameter $11.7$ mm ($\simeq 10$ grains) onto
the duralumin plate; the latter had a diameter larger than that of the
final heap. For wedge-shaped piles the sand emptied from a
wedge-shaped funnel with an outlet that was $11.7$ mm in the short
direction and $20$ cm in the long direction. The dimensions of the
supporting surface and the bottom of the wedge-shaped pile were $20$
cm X $26$ cm. Boundaries consisting of two Plexiglas walls $2.0$ cm
thick and taller than the peak of the pile preserved the wedge-shape
as the pile formed; the remaining two sides, parallel to the long
direction of the wedge, were open. The sensor was placed halfway
between the supporting walls, and at various distances from the
centerline of the heap. During the experiments we measured the volume
of the known mass of granular material forming the pile to determine
its average volume fraction, $\rho$.

{\em The raining procedure:} The second construction method was
designed to build up a pile in which the stress chains, and hence the
principal stress directions were more nearly in the vertical
direction. The containers from which the sand was poured had
cross-sectional dimensions slightly larger than the platform on which
the heap formed.  The bottom of these containers were wire meshes with
$0.40$ cm diameter holes.  To form the heaps, the containers were
filled while resting on the platform; they were then raised slowly
above the platform, allowing a steady rain of sand onto the heap.
Excess sand, at angles greater than the angle of repose, was allowed
to avalanche off the platform. For this procedure, the base platform
had the same size as the bottom of the final pile. The final mass of
sand and the pile volume was measured at the end of the procedure. For
conical piles we used a cylindrical container and a supporting
platform of diameter $26$ cm ($236$ grain diameters). For the
wedge-shaped piles, we used a rectangular box with dimensions $20$ cm
X $26$ cm; the platform was identical to the one used in the
localized-source procedure.

Pressure profiles and photographs of the final conical and
wedge-shaped piles are shown in Figs.~\ref{fig:cone} and
center of the heap is scaled by $R$, where $R$ is the pile radius for
conical heaps, and the distance from the center axis for wedge-shaped
heaps. The pressure is scaled by the hydrostatic pressure, $\rho
gH$. The bars represent the standard deviation of several {\it
independent} runs, not experimental error, which is about $0.25\%$.

The entire weight of each pile was integrated by curve-fitting the
profile and integrating over the base area. This calculation is then
compared with the known weight of the pile. For conical piles and
localized-source wedge-shaped piles the error between both
measurements is about $1.5\%$ or less. For the raining procedure
applied to wedge-shaped piles, we observe a discrepancy as large as $
8\%$. This relatively large ``missing mass'' for the wedge-shaped pile
may be caused by a screening effects of the walls which support some
of the weight.

Data for the conical piles created by the localized-source method show
a clear pressure minimum at $r/R = 0$.  A maximum in the stress of
$\sim 0.6\rho gH$, occurs at a position $r/R \cong 0.3$ which agrees
reasonably well with previous conical pile data
\cite{smid,brockbank}. The dimensionless stress at $r/R = 0$, $0.3\rho
gH$, is $\sim 50\%$ lower than the maximum stress. Experiments
performed with a fixed height funnel show a larger pressure difference
between the maximum at $r/R = 0$ and the value at $r/R = 0$.  This
suggests that the particles pack differently with different deposition
energies. The pressure is clearly largest near or at $r = 0$ in the
case of the raining procedure with maximum value $0.6\rho gH$.

A dip does not occur in the profiles of the heaps created by the
raining method. Rather, there is a peak pressure
of about $0.6$ at $r/R = 0$, and a steady drop in the pressure moving out
towards the edge of the pile.

For the wedge-shaped piles we find results qualitatively identical to
conical piles.  For the raining procedure, the stress profile shows no
indication of a central dip.  However, for the localized-source case
there is a clear minimum at $r/R = 0$. The value of this dip
is notably smaller than for the analogous conical heap, i.e. only
$15\%$ lower than the maximum stress, rather than $50\%$ lower.  The
pressure at the center is about $0.65\rho gH$ . The maximum in the
stress occurs at $r/R \cong 0.25$ with a value of about $0.75\rho gH$.
While the dip is smaller than the conical pile case, there is a
definite variation in the shapes of the profiles which indicates a
difference in the stress structure caused by the deposition
process. 

An important question concerns the dependence of the stress profile on
the heap size.  Earlier experiments\cite{jotaki,smid} suggested that
the size and relative position of the stress maximum may vary with the
size of the pile.  Alternatively, Radjai\cite{radjaib} has suggested
that the relative size of the funnel opening to the size of the heap
may be important.

\begin{figure}[]
\psfig{file=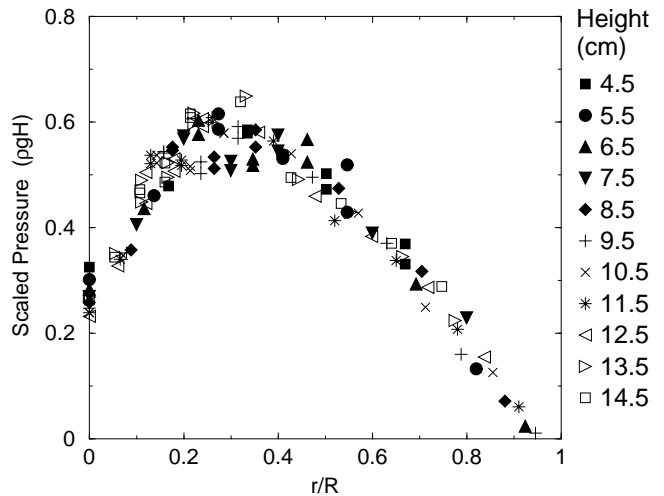,width=3.375in}
\caption{Normal stress profiles for different pile heights $H$ in localized
source experiments for a conical pile. The distance from the center of the
profiles is normalized by the pile radius $R$ and stresses are normalized by 
$\rho gH$.}
\label{fig:scaling}
\end{figure}

We have investigated the issue of heap size by considering conical
piles built with the localized source procedure. Specifically, we
obtained data for for heap heights spanning $4.5$ cm to $14.0$ cm by
simply stopping the filling process at various stages to obtain stress
data. This variation by $\sim 3$ in the maximum height of the piles
corresponds to a variation of $\sim 30$ in the mass, and hence the
peak stress.  The resulting data are displayed in
Fig.~\ref{fig:scaling}. While there is some scatter in the results,
the normalized profiles collapse surprisingly well. The peak occurs
consistently at $r/R =0.3$, and the stress at $r = 0$ is consistently
$\sim 50\%$ of the peak stress. This finding disagrees with earlier
studies by Jotaki et al.\cite{jotaki}, who also examine conical piles formed
by pouring from funnels.  These authors found that that the larger
piles had deeper dips in the stress at the center. The difference
between this data and ours is that Jotaki et al use a fixed funnel
height for a given heap height. Larger piles were formed by setting
the funnel progressively higher. Material dropped from the larger
heights had more energy than for smaller heights.  The height
dependence observed by Jotaki et al may be explained by a density
differences in the packings, with a corresponding height dependence of
the scaled stresses. In experiments where we fixed the funnel height
at a height $z > H$ above the base, we found that the stress dips were
deeper than for the experiments where we gradually raised the funnel.

To conclude, we have shown that the construction history affects the
pressure distribution at the bottom of a sand pile on a rigid
base. These experiments were conducted for conical and wedge shaped
piles. We observed the existence of a pressure dip at the center of a
sand pile if the filling procedure corresponded to a
localized-source. We found that the pressure profile scaled linearly
with the pile height, within the experimental scatter. It seems likely
that the progressive formation of the pile by successive small
avalanches leads to the occurrence of a pressure dip.  In the case of a
more uniformly vertical filling via a raining procedure, the dip
disappears. A localized-source procedure with a fixed pouring height
tends to produce a height dependent stress profile (with a dip). We
have shown that the dip in these experiments cannot be caused by a
deformation of the base. If small deflections of the base (order
$10^{-5}$) were an issue then, that effect should appear in both the
localized-source and raining procedures, and would also prevent the
collapse of the data for different heap heights..

A heuristic explanation of the mechanism producing the dip is that the
flow of particles during the localized-source procedure forms stress
chains oriented preferentially in the direction of the slope (c.f. Fig
1.). These chains form arches which shield the center from some of the
weight, thereby forming the dip. These effects agree qualitatively
with the explanations of Witmmer et al. Explanations for the magnitude
and of the dip its variation with geometry and are still lacking.  We
will present additional details and a more extensive comparison to
theory elsewhere.

{\bf Acknowledgments} We appreciate useful interactions with and comments
from Shigeyuki Tajima. The work of DH and RPB was supported by the National
Science Foundation under Grants DMR-9802602, and DMS-9803305, and by NASA
under Grant NAG3-1917. This work is also supported by a grant P.I.C.S.-563 
from the CNRS.\ Two of us (E.C. and L.V.) acknowledge the efficient
technical assistance and the pertinent advices of J.Lanuza, P.Lepert and
J.Servais.


\begin{references}
\bibitem{reviews}  For a broad review see H.~M. Jaeger and S.~R. Nagel and
R.~P. Behringer, {\em Physics Today}, {\bf 49}, 32 (1996) and {\em Rev. Mod.
Phys.} {\bf 68}, 1259 (1996); {\em Physics of Granular Media}, D. Bideau and
J. Dodds, eds. Les Houches Series, Nova (1991); {\em Granular Matter: An
Interdisciplinary Approach}, A. Mehta, Ed. Springer, NY (1994); R.~P.
Behringer, Nonlinear Science Today, {\bf 3}, 1 (1993).

\bibitem{chains}  P. Dantu, G\'{e}otechnique, {\bf 18}, 50 (1968); A.
Drescher and G. De Josselin De Jong, J.Mech. Phys. Solids, {\bf 20}, 337
(1972); T. Travers et al. J. Phys. A. {\bf 19}, L1033.

\bibitem{radjai}  F.Radjai, D.Wolf, M.Jean and J.J.Moreau, Phys. Rev.Lett. 
{\bf 80}, 61 (1998) and references therein.

\bibitem{howell}  D.Howell and R.P.Behringer, in {\it Powder and Grains 97},
p 337, ed. by Behringer and Jenkins, (Balkema, Rotterdam 1997).

\bibitem{ouagenouni}  S.Ouagenouni and J.N. Roux , Europhys.Lett.{\bf 39,}
117 (1997)

\bibitem{eloy}  C.Eloy and E.Cl\'{e}ment, J.de Phys.I (France) {\bf 7}, 1541
(1997).

\bibitem{bouchaud}  J.-P Bouchaud, M.E. Cates, and P. Claudin, {\it J.
Physique I} {\bf 5}, 639 (1995).

\bibitem{watson}  A. Watson, {\it Science} {\bf 273}, 579 (1996).

\bibitem{wittmer}  J. Wittmer, P. Claudin, M.E. Cates, and J.-P. Bouchaud, 
{\it Nature} {\bf 382}, 336 (1996).

\bibitem{wittmer2}  J. Wittmer, M.E. Cates, and P. Claudin, {\it J. Physique
I} {\bf 7}, 39 (1997).

\bibitem{savage} for an extensive discussion of past experiments as
well as a discussion of soil mechanics models see: S. B. Savage, {\it
Proceeding of the Third International Conference on Powders \&
Grains}, A.A. Balkema, Rotterdam, Netherlands (1997), and references
therein.

\bibitem{goddard}  F. Cantelaube and J.D. Goddard, {\it Proceeding of the
Third International Conference on Powders \& Grains}, A.A. Balkema,
Rotterdam, Netherlands (1997).

\bibitem{hummel}  F.H. Hummel and E.J. Finnan,{\it Proc. Instn. Civil Eng.} 
{\bf 212}, 369 (1921).

\bibitem{jotaki}  T. Jotaki and R. Moriyama, {\it J. Soc. Poder Technol. Jpn.%
} {\bf 60}, 184 (1979).

\bibitem{smid}  J. Smid and J. Novosad, {\it Proc. of 1981 Powtech
Conferenc, Ind. CHem Eng. Symp.} {\bf 63}, D3/V/1-D3/V12 (1981).

\bibitem{brockbank}  R. Brockbank , J.M. Huntley, and R. Ball, {\it J.
Physique I France} {\bf 10}, 1521 (1997).

\bibitem{brown}  R.L.Brown and J.C.Richards, {\it Principle of Powder
Mechanics}, Pergamon (1970).

\bibitem{vanel}  L.Vanel, E.Clement , Eur.Phys.J. B, {\it to be published}
(1999);{\it \ preprint} Cond-Mat/9904090.

\bibitem{savage98}  S.B. Savage, in {\it Dry Granular Media}, eds. H.~J.
Herrmann, S. Luding, J.~P. Hovi, NATO ASI series, Kluver, Amsterdam
(1998).

\bibitem{device} A commercial pressure measuring device, the Micro-Epsilon
Messtechnik Series 610, was used to measure the pressure.

\bibitem{huang}  J.H.S. Huang and S.B. Savage, Wall Stresses Developed by
Granular material in Axisymmteric Bins Dept. of Civil Engineering \& Applied
Mechanics, McGill University, FML-TR 70-1, 130 (1970).

\bibitem{lee}  I.F. Lee and J.R. Herrington, {\it 1st Aust.-N.Z. Conf.
Geomech.} {\bf 1}, 291 (1971).

\bibitem{trollope}  D.H. Trollope and B.C. Burman, {\it Geotechnique} {\bf 30%
}, 137 (1980).

\bibitem{radjaib}  F.Radjai, {\it private communication.}
\end{references}
\end{document}